\documentstyle{elsart}
\input psfig.tex

\newcommand{\widtheps}{12cm}

\newcommand{\diag}{{\mathrm{diag}}}
\newcommand{\tr}{{\mathrm{tr}}}

\begin{document}

\begin{frontmatter}
\title{Two photons decay of the glueball and scalar isoscalar
 mesons in a scaled NJL model}

\author[madrid]{M. Jaminon\thanksref{onleave}},
\author[liege]{B. Van den Bossche\thanksref{iisn}}
\address[madrid]{Instituto de Fisica Teorica, Universidad
Autonoma de Madrid, Spain}
\address[liege]{Universit\'e de Li\`ege, Institut de Physique 
B5, Sart Tilman, B-4000~Li\`ege~1, Belgium}
\thanks[onleave]{On leave of absence from
Universit\'e de Li\`ege, Institut de Physique B5, 
Sart Tilman, B-4000 Li\`ege 1, Belgium}
\thanks[iisn]{Supported by the Institut Interuniversitaire 
des Sciences Nucl\'eaires de Belgique}

\begin{abstract} We use a modified version of the Nambu--Jona-Lasinio
model which implements the QCD trace anomaly to calculate the two photons
decay width of the glueball ($f_{0}(1500)$)  and of the two scalar
mesons ($f_{0}(1370)$, $f_{J=0}(1710)$) to which it is mixed.
We
investigate the effect of this mixing over the coupling constants of
the $f_0$ states to the quarks.
\end{abstract}       

\begin{keyword} NJL, Scalar mesons, Glueball, Two photons decay
\PACS  12.39.Fe, 12.39.Mk, 14.40.-n
\end{keyword}

\end{frontmatter}

\section{Introduction}
\label{section1}
The scalar resonance $f_{0}(1500)$ which has been discovered at LEAR by
the Crystal Barrel Collaboration in the reaction $p\overline{p}$ (at
rest) $\rightarrow 3{\pi}^{0},\eta \eta {\pi}^{0}$ \cite{1} and $\eta
\eta\ '{\pi }^{0}$ \cite{2}, was immediately considered as a good
candidate for the scalar glueball. Indeed, the observed mass
((1520$\pm$25) MeV in \cite{1} and (1545$\pm$25) MeV in \cite{2}) was
near the value predicted by the lattice-QCD calculations \cite{3} of
that time. It is also in agreement with the analysis of recent data on the 
decay
$f_{0}(1500)\rightarrow K\overline{K}$ \cite{3bis} which however requires a 
mixture
between the glueball and the neighbouring $q\overline{q}$ states \cite{4}. 
Another
signal that strengthens this identification lies in the fact that, in
the radiative decay $J/\psi \rightarrow \gamma+4\pi$, a sharp peak is
seen in the 4$\pi$ spectrum, having quantum numbers ${0}^{++}$ and
mass and width also compatible with $f_{0}(1500)$ \cite{5}.

The mixing between the glueball and the $q\overline{q}$ states has also been 
noted
by Sexton et al. \cite{4bis}. They however identify the nearly glueball state 
with
$f_{J}(1710)$  whose mixing with $f_{0}(1370)$ and $f_{0}(1500)$ requires 
that
 $J = 0$ \cite{4}, \cite{6}.

In Refs. \cite{6}, \cite{7}, it has been claimed that good tests of
the glue and $q\overline{q}$ contents of the glueball 
 would consist in studying its partial strong decay widths
into $\pi\pi,
K\overline{K},
\eta\eta', \eta\eta$. We plan to calculate them
in a forthcoming paper \cite{9}.

In the present paper, we follow the idea of Close et al. \cite{4} who estimate 
the two photons decay widths of the glueball and of the mesons to which it is
mixed.
 Emission of photons is surely a good test to investigate the glueball
or $q\overline{q}$-nature of the $f_{0}$ states since gluons do
not couple to photons. 
The authors suggest a mixing between $f_{0}(1370)$, $f_{0}(1500)$ and 
$f_{0}(1710)$ for
which they study two different schemes, according to the preceding remarks.
Firstly, they consider the case where $f_{0}(1500)$ exhibits large contents 
in
glue and in $q\overline{q}$ ($u$ and $s$) excitations while $f_{0}(1710)$ has its 
largest
component in the  $s\overline{s}$ chanel. This allows to reproduce  the
$f_{0}(1500)\rightarrow K\overline{K}$ data \cite{3bis}. The second scheme 
consists
in considering the glueball lying above the $s\overline{s}$ member of the 
nonet
\cite{4bis}:  $f_{0}(1710)$ can then be the glueball while $f_{0}(1500)$ is 
mainly
a $s\overline{s}$ excitation. Within these two schemes, Close et al. \cite{4} 
have
estimated the relative strength of the $2\gamma$ widths for the three $f_0$ 
states:

\[ f_0(1370):f_0(1500):f_0(1710)\approx12:1:3,\hspace{10mm}(\mbox{scheme 1})\]
\[f_0(1370):f_0(1500):f_0(1710)\approx13:0.2:3.\hspace{7mm}(\mbox{scheme 2})\]

 Whatever the mixing, the decay width of $f_{0}(1500)$ is always the
smallest. Experimental estimation of the $2\gamma$ widths  would then be a 
good
test of the general idea of $q\overline{q}$  and glue mixing. Indeed, if the
width of the $f_{0}(1500)$ was found to be larger than one of the others, the 
idea
of mixing would be destroyed. Note that the decay
$f_{0}(1370)\rightarrow 2\gamma$  is 
the only one that has been measured experimentally : 
$\Gamma_{f_0(1370)}=5.4\pm2.3$
 KeV \cite{14}.

\begin{figure}[h]
\vbox{\psfig{file=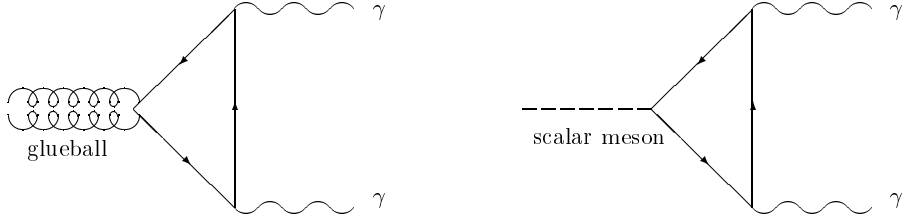,width=\widtheps}}
\caption{Schematic $2\gamma$ decay of the glueball and of a scalar state.}
\label{fig1}
\end{figure}

The model we use to calculate the two photons decay widths 
is a modified version of the SU(3)
NJL model \cite{10} which implements the trace
 anomaly of QCD \cite{11}. It is called the
scaled NJL model. It entails the introduction of a scalar field $\chi$ whose 
mean
value $\chi_0$ can be identified with the vacuum gluon condensate
\cite{12}. This $\chi$ field couples to the quarks so that our model
can describe the processes schematically shown in Fig.~\ref{fig1}. In our model, it is 
the vacuum gluon condensate that fixes the mixing
\cite{13}: the smaller $\chi_0$, the larger the mixing. 

However, in the 
domain of
$\chi_0$ which allows to roughly reproduce the meson masses, $f_{0}(1710)$
remains mainly a $s\overline{s}$ excitation. Its coupling to the up quark
is weak. In contrast, the strength of the coupling to the $u$ or $s$ quark
is roughly of the same order of magnitude for $f_{0}(1370)$ and for 
$f_{0}(1500)$. In our model, the three states have glue content. 
$f_{0}(1500)$ is the one which
yields to pure glueball if no mixing and it is the state
that always has the largest glue content even if that of $f_{0}(1370)$
can become important. We then work in a scheme similar to the scheme 1 
presented above.

 The parameters of the model
are fixed to reproduce the pion  and kaon masses, the pion decay constant and 
the
masses of
$a_0(1450)$
$[{u\overline{d},\overline{u}d,
{1 \over \sqrt
{2}}(u\overline{u}-d\overline{d})}]$ and $f_{0}(1500)$. The mass of the 
largest member of the nonet ($f_{0}(1710)$)
will then vary as well as that of $f_{0}(1370)$. The latter is very sensitive 
to the value of the vacuum gluon condensate while the former keeps a value 
not far away from $1710$ MeV, whatever $\chi_{0}$ . 

For completeness, let us add that the $2\gamma$ decay of the scalar mesons 
has also
been  studied in Ref.  \cite{14bis} within a relativistic quark
model with linear confinement and instanton-induced
interaction. $f_{0}(1500)$ and $f_{0}(980)$
 are there considered as mixing between $u\overline{u}$ and
 $s\overline{s}$ states. Another relativistic treatment using a OGE model
has been presented in Ref.  \cite{MU}.

Our paper is organized as follows. Section~\ref{section2} recalls the useful tools
of the scaled NJL model. Section~\ref{section3} is devoted to the derivation of the
decay widths of the mesons and of the glueball. We insist on the fact
that the mixing between the three fields generating the glueball and
the scalar mesons modifies the transition amplitudes. For instance,
$f_{0}(1370)$ can decay into $2\gamma$ through the production of a
$s\overline{s}$ pair. The estimation of the widths assumess  the
knowledge of the masses of the mesons and the glueball as well as
their coupling constants to the quarks. These quantities are defined
in Section~\ref{section4}. We discuss our results in Section~\ref{section5}. Finally, Section~\ref{section6}
gives our conclusions.

\section{The model}
\label{section2}
The model used in this paper is  described in Refs.  \cite{10,15}
under the name of ''A-scaling model''. We  recall here some of the
useful tools for the understanding of the present work. We start from
the vacuum SU(3) effective Lagrangian:
\begin{eqnarray}
{\cal L}_{eff}(q,\bar{q}, \chi)&=& \bar{q}(-i{\partial
}_{\mu}\gamma_{\mu}+m)q
\nonumber\\
&&\mbox{}+\frac{1}{2a^2\chi^2}\sum_{i=0}^{8}[(\bar{q}\frac{(\lambda^i)_{F}}{2}q)^{2}+(\bar{q}i\gamma_5\frac{((\lambda^{i})_{F}}{2}q)^{2}]+{\cal L}_{\chi }
\label{Lqq}
\end{eqnarray}
which yields to the bosonized effective Euclidean action
\begin{eqnarray}
I_{eff}(\varphi, \chi)&=&-{\rm Tr}_{\Lambda \chi }\ln(-i{\partial
}_{\mu}\gamma_{\mu}+m+\Gamma_{a}\varphi_{a})\nonumber\\
&&\hspace{3cm}\mbox{}+\int_{}^{}{d}^{4}x
\frac{a^{2}\chi^{2}}{2}\varphi_{a}\varphi
_{a}+\int_{}^{}{d}^{4}x{\cal L}_{\chi }
\label{Ieff}
\end{eqnarray}
once one has integrated out the quark degrees
of freedom. The meson fields write:
\begin{equation}
\varphi_{a}=(\sigma_{a},\pi_{a}),\hspace{15mm} \Gamma
_{a}=(\lambda_{a},i\gamma _{5}\lambda_{a}),\hspace{15mm}a=0,...,8
\label{champ}
\end{equation}
where the $\lambda _{a}$ are the usual Gell-Mann matrices
with $\lambda _{0}= {\sqrt{2/3} \bf{1}}$. We choose to work in the isospin
symmetry limit and the quantity $\it m$ stands for the diagonal matrix
$\diag(m_{u},m_{u},m_{s})$. The trace anomaly of QCD is
modelized using a scalar dilaton field $\chi$ that is intimately
related to the gluon condensate $\chi\propto{\langle 
G^2_{\mu\nu}\rangle}^{1/2}$
 {\cite{10}}:
\begin{equation}
{\cal L_{\chi}}=\frac{1}{2}(\partial_{\mu}\chi)^{2}+\frac{1}{16}b^{2}
(\chi^{4}\ln \frac{\chi^{4}}{\chi^{4}_{G}}-(\chi^{4}-\chi^{4}_{G})).
\label{lagci}
\end{equation}
Since we are only interested in the scalar sector, the axial anomaly
which would give the {$\eta$-$\eta$'}  mass difference is not considered
here. Our model contains six parameters: the current quark masses
 ($m_{u}, m_{s}$), the strengths ($a^{2}, b^{2}$), the gluon parameter 
 $\chi_{G}$ and the cut-off $\Lambda$ introduced to regularize the
 diverging quark loop. Four of these parameters are
adjusted to reproduce the pion mass ($m_{\pi}$), the weak pion decay constant 
($f_{\pi}$), the kaon mass ($m_{K}$) and that of {$f_{0}$}(1500). We
then have two free parameters that we choose to be the vacuum gluon
condensate $\chi_0$ (related to $\chi_G$ {\cite{15}}) and the constituent
up quark mass $M_u$. The latter is connected to the strength $a^{2}$
using the usual gap equation {\cite{10}} which minimizes the action. Note
that  the strange constituent quark mass $M_s$ and $\chi_0$ obey
similar stationary conditions. $M_u$ will have a large value  ($725$
MeV) in order to reproduce
$a_{0}(1450)$ \cite{6} and $\chi_0$ will remain free.

\section{The decay widths}
\label{section3}
The widths for the two photons decay of the scalar isoscalar mesons are
given by
\begin{eqnarray}
\Gamma_i&=&{1 \over 2}{1 \over 2{m}_{i}}{1 \over (2\pi
{)}^{2}}\int\int{{d}^{3}{k}_{1} \over
2\left|{{\vec{k}}_{1}}\right|}{{d}^{3}{k}_{2} \over
2\left|{{\vec{k}}_{2}}\right|}{\left|{{\cal T}_{i}}\right|}^{2}
\delta(Q-{k}_{1}-{k}_{2})\nonumber\\
&=&{1 \over 32\pi }{1 \over {m}_{i}}
{\left|{{\cal T}_{i}}\right|}^{2}
\label{eq5}
\end{eqnarray}
for $i=f_{0}(1370), f_{0}(1710)$ and $f_{0}(1500)$. In Eq.~(\ref{eq5}), $m_i$ represents the mass of the 
meson, $Q$ its quadrimomentum while $k_1$ and $k_2$ stand for the quadrimomenta of the emerging photons. The mesons are
assumed to be at rest. In a pure NLJ, the transition amplitude for
$f_{0}(1500)$, identified with the glueball, is identically zero;
${\Gamma }_{{f}_{0}(1370)}$ only involves the propagator of the up quark
(in the limit of isospin symmetry) while ${\Gamma }_{{f}_{0}(1710)}$  is
expressed in terms of the strange quark propagator. 

\begin{figure}[h]
\vbox{\psfig{file=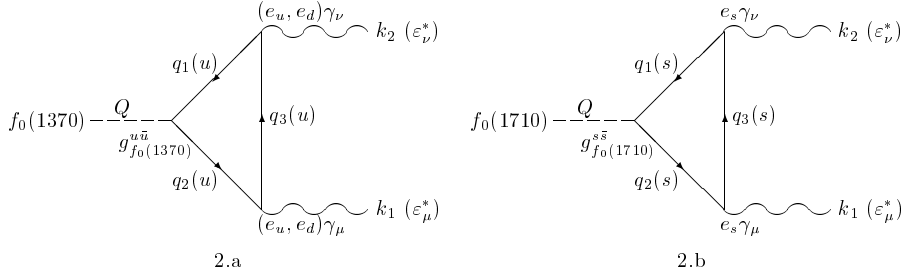,width=\widtheps}}
\caption{Feynman direct diagrams for $2\gamma$ decay of $f_0(1370)$ (a)
and  of $f_0(1710)$ (b).}
\label{fig2}
\end{figure}

The direct diagram
(Fig.~\ref{fig2}.a) writes
\begin{equation}
{\cal T}_{{f}_{0}(1370)}={N}_{c}({e}_{u}^{2}+{e}_{d}^{2})\
{g}_{{f}_{0}(1370)}^{u\overline{u}}\ {\varepsilon }_{\mu
}^{*}({k}_{1}){\varepsilon }_{\nu }^{*}({k}_{2})U
\label{eq6}
\end{equation}
with
\begin{eqnarray}
U&=&\int{d^4q_3 \over (2\pi {)}^{
4}}\tr_D\left\{{({q\!\!\!/}_2+{M}_{u}){\gamma
}_{\mu}({q\!\!\!/}_3+{M}_{u}){\gamma
}_{\nu}({q\!\!\!/}_1+{M}_{u})}\right\}\nonumber\\ &\times&{1 \over
({q}_{1}^{2}+{M}_{u}^{2})({q}_{2}^{2}+{M}_{u}^{2})
({q}_{3}^{2}+{M}_{u}^{2})}
\label{eq7}
\end{eqnarray}
while the direct diagram (Fig.~\ref{fig2}.b) has the similar form:
\begin{equation}
{\cal
T}_{{f}_{0}(1710)}={N}_{c}{e}_{s}^{2}
{g}_{{f}_{0}(1710)}^{s\overline{s}}
{\varepsilon }_{\mu
}^{*}({k}_{1}){\varepsilon }_{\nu }^{*}({k}_{2})S
\label{eq8}
\end{equation}
with
\begin{eqnarray}
S&=&\int{d^4q_3 \over (2\pi {)}^{t
4}}\tr_{D}\left\{{({q\!\!\!/}_2+{M}_{s}){\gamma
}_{\mu}({q\!\!\!/}_3+{M}_{s}){\gamma
}_{\nu}({q\!\!\!/}_1+{M}_{s})}\right\}\nonumber\\ &\times&{1 \over
({q}_{1}^{2}+{M}_{s}^{2})({q}_{2}^{2}+{M}_{s}^{2})
({q}_{3}^{2}+{M}_{s}^{2})}.
\label{eq9}
\end{eqnarray}

In the scaled NLJ, the mesons $f_{0}(1370)$ and $f_{0}(1710)$ are not
anymore pure $u\overline{u}$ or pure $s\overline{s}$ excitations.
Equations (\ref{eq6}) and (\ref{eq8}) have then to be modified in order to 
take
these new effects into account:
\begin{eqnarray}
{\cal
T}_{{f}_{0}(1370)}&=&{N}_{c}({e}_{u}^{2}+{e}_{d}^{2}){g}_{{f}_
{0}(1370)}^{*u\overline{u}
}{\varepsilon }_{\mu }^{*}({k}_{1}){\varepsilon }_{\nu
}^{*}({k}_{2})U \nonumber\\
&&\hspace{2cm}\mbox{}+{N}_{c}{e}_{s}^{2}{g}_{{f}_{0}(1370)}^{*s\overline{s}
}{\varepsilon
}_{\mu }^{*}({k}_{1}){\varepsilon }_{\nu }^{*}({k}_{2})S
\label{eq10}
\end{eqnarray}
\begin{eqnarray}
{\cal
T}_{{f}_{0}(1710)}&=&{N}_{c}{e}_{s}^{2}{g}_{{f}_{0}(1710)}^{*s\overline{s}
}{\varepsilon }_{\mu }^{*}({k}_{1}){\varepsilon }_{\nu
}^{*}({k}_{2})S \nonumber\\
&&\hspace{2cm}\mbox{}+{N}_{c}({e}_{u}^{2}+{e}_{d}^{2}){g}_{{f}_{0}(1710)}^{*u\overline{u}
}{\varepsilon }_{\mu }^{*}({k}_{1}){\varepsilon }_{\nu
}^{*}({k}_{2})U.
\label{eq11}
\end{eqnarray}
The quark meson coupling constants
${g}_{i}^{*u\overline{u}}$, ${g}_{i}^{*s\overline{s}}$, ($i=f_0(1370)$,
 $f_0(1710)$) differ from the ones introduced in Eqs. (\ref{eq6}) and
(\ref{eq8}). Their analytical expressions will be established in Section.~\ref{section4}.   
 
Having some content in $q\overline{q}$ excitations the
glueball $f_{0}(1500)$  can now decay:
\begin{eqnarray}
{\cal
T}_{{f}_{0}(1500)}&=&{N}_{c}({e}_{u}^{2}+{e}_{d}^{2}){g}_{{f}_{0}(1500)
}^{*u\overline{u}\
}{\varepsilon }_{\mu }^{*}({k}_{1}){\varepsilon }_{\nu
}^{*}({k}_{2})U \nonumber\\
&&\hspace{2cm}\mbox{}+{N}_{c}{e}_{s}^{2}{g}_{{f}_{0}(1500)}^{*s\overline{s}
}{\varepsilon }_{\mu }^{*}({k}_{1}){\varepsilon }_{\nu
}^{*}({k}_{2})S.
\label{eq12}
\end{eqnarray}
Summing over the final polarization of the photon and using
$Q=k_1+k_2$, $q_1=q-k_2$, $q_2=q+k_1$, $q_3=q$, one gets :
\begin{eqnarray}
{\left|{{\cal T}_{{f}_{0}(1370)}}\right|}^{2}&=&{12800 \over 9}{\pi
}^{2}{\alpha
}^{2}\bigg|{({m}_{{f}_{0}(1370)}^{2}-4{M}_{u}^{2})
{M}_{u}{g}_{{f}_{0}(1370)}^{*u\overline{u}}{J}_{u}
(-{m}_{{f}_{0}(1370)}^{2})}\bigg. \nonumber\\
&+&{\bigg.{{1 \over
5}({m}_{{f}_{0}(1370)}^{2}-4{M}_{s}^{2}){M}_{s}{g}_{{f}_{0}
(1370)}^{*s\overline{s}}{J}_{s}(-{m}_{{f}_{0}
(1370)}^{2})}\bigg|}^{2}
\label{eq13}
\end{eqnarray}

\begin{eqnarray}
{\left|{{\cal T}_{{f}_{0}(1710)}}\right|}^{2}&=&{12800 \over 9}{\pi
}^{2}{\alpha
}^{2}\bigg|{({m}_{{f}_{0}(1710)}^{2}-4{M}_{u}^{2})
{M}_{u}{g}_{{f}_{0}(1710)}^{*u\overline{u}}{J}_{u}
(-{m}_{{f}_{0}(1710)}^{2})}\bigg. \nonumber\\
&+&{\bigg.{{1 \over
5}({m}_{{f}_{0}(1710)}^{2}-4{M}_{s}^{2}){M}_{s}{g}_{{f}_{0}
(1710)}^{*s\overline{s}}{J}_{s}(-{m}_{{f}_{0}
(1710)}^{2})}\bigg|}^{2}
\label{eq14}
\end{eqnarray}

\begin{eqnarray}
{\left|{{\cal T}_{{f}_{0}(1500)}}\right|}^{2}&=&{12800 \over 9}{\pi
}^{2}{\alpha
}^{2}\bigg|{({m}_{{f}_{0}(1500)}^{2}-4{M}_{u}^{2})
{M}_{u}{g}_{{f}_{0}(1500)}^{*u\overline{u}}{J}_{u}
(-{m}_{{f}_{0}(1500)}^{2})}\bigg. \nonumber\\
&+&{\bigg.{{1 \over
5}({m}_{{f}_{0}(1500)}^{2}-4{M}_{s}^{2}){M}_{s}{g}_{{f}_{0}
(1500)}^{*s\overline{s}}{J}_{s}(-{m}_{{f}_{0}
(1500)}^{2})}\bigg|}^{2}
\label{eq15}
\end{eqnarray}
with
\begin{equation}
{J}_{i}(Q)=\int_{}^{}{{d}^{4}q \over (2\pi {)}^{4}}{1 \over
\left[{(q+{k}_{1}{)}^{2}+{M}_{i}^{2}}\right]
\left[{(q-{k}_{2}{)}^{2}+{M}_{i}^{2}}\right]\left
[{{q}^{2}+{M}_{i}^{2}}\right]}
\label{eq16}
\end{equation}
and $\alpha ={e}^{2}/4\pi$ the fine structure
constant\footnote{Detailed calculations for ${\pi}_{0}\rightarrow
\gamma\gamma$ can be found in Ref.  {\cite{15}}}. We will 
work with a large constituent up quark mass,
 $M_u$ = 725 MeV, in order to reproduce $a_{0}(1450)$.
However, even with such a large mass, the mesons 
$f_{0}(1710)$ and the glueball (1500) always lie above
 the unphysical pair
$u\overline{u}$ creation threshold $2M_u$. In the same way, 
$f_{0}(1710)$, which aquires a mass larger than $2M_s=1728$ MeV,
 is always above the $s\overline{s}$ threshold. The function $J_u$ should
then be calculated in the complex plan:

\begin{equation}
{J}_{u}\rightarrow {\left.{{J}_{u}
({Q}^{2}-i{0}^{+})}\right|}_{{Q}^{2}=-({m}_{i}-{i{\Gamma
}_{i} \over 2}{)}^{2}}, 
\label{eq17}
\end{equation}
where $\Gamma_i$ is the width of the meson for its $u\overline{u}$
decay and with a similar expression for $J_s$. This width
 as well as the mass $m_i$ can be calculated
from a Dyson-Schwinger equation (see below) which amounts to two
coupled equations for $\Gamma_i$ and $m_i$. The latter cannot be
decoupled without a suitable and model-dependent description. In the
same way, expression (\ref{eq17}) is here approximated by
$\left|J_u(-{m_i}^2-i0^+)\right|$. Note that the results are not
significantly
affected by the prescription used above the threshold. However,
 it would worth using a confining model to get rid of the unphysical
decay channel. Indeed, Celenza et al. {\cite{BROO}} have shown that
 a modification of the 
scalar vertex by a confining nonlocal interaction increases the mass
 of the ${\sigma}$ of $\approx 250$ MeV. This feature could surely modify 
quantitatively the results we present here.

\section{Meson masses and coupling constants}
\label{section4}
Equations (\ref{eq13}-\ref{eq15}) involve the masses of the mesons as well as 
their
coupling constants to the quarks $u$ and $s$. We now show how to calculate
them, insisting on the way the coupling constants are modified due to
the coupling between the various fields. The effective action (\ref{Ieff}) 
can be expanded up to
second order in the fluctuating parts of the meson fields $({\tilde{\sigma
}}_{0},{\tilde{\sigma }}_{8})$ and of the dilaton field $\tilde
{\chi}$ :
\begin{equation}
{I}^{(2)}({\sigma }_{0},{\sigma }_{8},\chi )={1 \over 2\beta \Omega
}\sum\limits_{q}^{} ({\tilde{\sigma }}_{0,q},{\tilde{\sigma
}}_{8,q},{\tilde{\chi }}_{q}){S}^{-1}\left({\matrix{{\tilde{\sigma
}}_{0,-q}\cr {\tilde{\sigma }}_{8,-q}\cr {\tilde{\chi
}}_{-q}\cr}}\right).
\label{eq18}
\end{equation}
$S^{-1}$ is a 3x3 matrix with 9 nonvanishing elements that can be read
in Ref.  \cite{10}. It can be diagonalized, yielding to
\begin{equation}
{I}^{(2)}({\phi }_{1},{\phi }_{2},{\phi }_{3})={1 \over 2\beta \Omega
}\sum\limits_{ q}^{} \sum\limits_{i}^{} {\widetilde{\phi }}_{
i,q}\diag\left[{{\Lambda }_{ii}({q}^{2})}\right]{\widetilde{\phi
}}_{i,-q}.
\label{eq19}
\end{equation}
with
\begin{equation}
\diag\left[{{\Lambda }_{ii}({q}^{2})}\right]=
{V}^{T}(q^{2}){S}^{-1}V(q^{2})
\label{eq19bis},
\end{equation}
and $V$ the $q^2$ dependent eigenvector matrix. Expression (\ref{eq19}) can 
be
approximated by expanding the elements of the diagonal matrix around their
respective zero, that is around the respective meson masses:
\begin{eqnarray}
{I}^{(2)}({\phi }_{1},{\phi }_{2},{\phi }_{3})\approx{1 \over 2\beta \Omega
}\sum\limits_{q}^{} \sum\limits_{i}^{} {\widetilde{\phi }}_{i,q}
\diag\bigg[{{\Lambda }_{ii}}\bigg.(-{m}_{i}^{2})\cr
\noalign{\medskip}
\bigg.+({q}^{2}+{m}_{i}^{2}){\partial }_{{q}^{2}}{\Lambda }_{ii}
{\bigg.{({q}^{2})}\bigg|}_{{q}^{2}=-{m}_{i}^{2}}\bigg]{\widetilde{\phi}}_{i,-
q}\cr
\nonumber \\ 
\noalign{\medskip}
={1 \over 2\beta \Omega }\sum\limits_{q}^{}
\sum\limits_{i}^{} {\widetilde{\phi }}_{i,q}{G}_{i}^{-2}
({q}^{2}+{m}_{i}^{2}){\widetilde{\phi }}_{i,-q}.
\label{eq20}
\end{eqnarray}
In Eq. (\ref{eq20}), we have used the fact that
\begin{equation}
\diag\left[{{\Lambda }_{ii}(-{m}_{i}^{2})}\right]=0
\label{eq21}
\end{equation}
and we have defined \cite{13} 
\begin{equation}
{G}_{i}^{-2}={\partial }_{{q}^{2}}\diag\left[{\Lambda
}_{ii}(-{m}_{i}^{2})\right]={\left[{{V}^{T}(-{m}_{i}^{2})({\partial
}_{{q}^{2}}{S}^{-1})V(-{m}_{i}^{2})}\right]}_{ii}
\label{eq22}
\end{equation}
where the eigenvector matrix $V$ has its first row calculated at
$-{m}_{1}^{2}=-{m}_{{f}_{0}(1370)}^{2}$, its second row at
$-{m}_{2}^{2}=-{m}_{{f}_{0}(1710)}^{2}$ and the last one at
$-{m}_{3}^{2}=-{m}_{{f}_{0}(1500)}^{2}$.
The new physical fields ${\widetilde{\phi }}_{i}^{{}^*}$ associated
 with the three $f_0$ states write:
\begin{equation}
{\widetilde{\phi }}_{i}^{{}^*}={G}_{i}^{-1}{\widetilde{\phi
}}_{i}={G}_{i}^{-1}\sum\limits_{ j}^{} {V}_{ij}^{-1}
(-{m}_{i}^{2}){\widetilde{\sigma }}_{j}
\label{eq24}
\end{equation}
with ${\widetilde{\sigma }}_{1}\equiv {\widetilde{\sigma }}_{0}
,{\widetilde{\sigma }}_{2}\equiv {\widetilde{\sigma }}_{8}
,{\widetilde{\sigma }}_{3}\equiv \widetilde{\chi}$.

The interaction term in the quark loop ($\ln$) between the quarks and the
 meson fields (see Eq.(\ref{Ieff})) reflects  an interaction term in the 
Lagrangian
of the form $\overline{q}({\sigma}_{0}{\lambda}_{0}+{\sigma}_{8}{\lambda
}_{8})q$. Using Eq. (\ref{eq24}), the latter becomes an interaction term 
between
the quarks and the ${\widetilde{\phi }}_{i}^{{}^*}$:
\begin{equation}
\overline{q}({\sigma }_{0}{\lambda }_{0}+{\sigma }_{8}{\lambda
}_{8})q=\overline{u}{A}_{1}u+\overline{d}{A}_{2}d+\overline{s}{A}_{3}
{s}
\label{eq25}
\end{equation}
with (there is no summation over the index i)
\begin{equation}
{A}_{i}=\sum\limits_{j=1,2,3}^{} {\widetilde{\phi }}_{j}^{^*}
\left({
({\lambda }_{0}{)}_{ii}{V}_{1j}+({\lambda
}_{8}{)}_{ii}{V}_{2j}}\right){G}_{j}\ \ \ \ \ i=1,2,3
\label{eq26}
\end{equation}
showing that the two mesons as well as the glueball are indeed coupled
to the three quarks with the respective coupling constants:
\begin{eqnarray}
{g}_{j}^{*u\overline{u}}={g}_{j}^{*d\overline{d}}&=&\left[{({\lambda
}_{0}{)}_{11}{V}_{1j}+({\lambda
}_{8})_{11}{V}_{2j}}\right]{G}_{j}
\label{eq27}
\\
g_j^{*s\overline{s}}&=&\left[{({\lambda
}_{0}{)}_{33}{V}_{1j}+({\lambda }_{8}{)}_{33}{V}_{2j}}\right]{G}_{j},\label{eq28}\\
&&\hspace{2cm} j=1({f}_{0}(1370)), 2({f}_{0}(1710)), 3({f}_{0}(1500)).
\nonumber
\end{eqnarray}
In a pure SU(3) NJL, without mixing with the dilaton, Eq. (\ref{eq26}) reduces 
to
\begin{equation}
{A}_{i}={\widetilde{\phi}}_{i}^{{}^*}{G}_{i}\ \ \ \ \ i=1,2
\label{eq29}
\end{equation}
so that Eq. (\ref{eq22}) defines the coupling constant of the mesons to the
quark and ${\widetilde{\phi}}_{1}^{{}^*}\
({\widetilde{\phi}}_{2}^{{}^*})$ is only coupled to the quark $u (s)$.

In the present approach, we will use 
Eqs. (\ref{eq21}), (\ref{eq27}), (\ref{eq28})
 to calculate the mass of the mesons and their respective coupling to the 
quarks.
Note that it is the nonvanishing value of
${g}_{f_0(1500)}^{*u\overline{u}}$ and ${g}_{f_0(1500)}^{*s\overline{s}}$ that
allows the  glueball decay $f_{0}(1500) \rightarrow \gamma \gamma$.

As mentioned above, some caution must be taken when the
$u\overline{u}$ or $s\overline{s}$ thresholds are reached.
 Due to the mixing with the
glueball, it is not possible to solve  analytically  $\diag\left[{{\Lambda }_{ii}({q}^{2})}\right]$ (see Eq. (\ref{eq21})). In
order to illustrate the threshold problem, we will drop here this mixing. The
diagonalization then yields:
\begin{equation}
{\Lambda
}_{11}^{{}^*}({q}^{2})=4{N}_{c}{F}_{u}(q^2)({q}^{2}+4{M}_{u}^{2})+{a}^{2}
{\chi
}_{0}^{2}{{m}_{u} \over {M}_{u}}
\label{eq30}
\end{equation}
\begin{equation}
{\Lambda
}_{22}^{{}^*}({q}^{2})=4{N}_{c}{F}_{s}(q^2)({q}^{2}+4{M}_{s}^{2})+{a}^{2
}{\chi
}_{0}^{2}{{m}_{s} \over {M}_{s}}
\label{eq31}
\end{equation}
with
\begin{equation}
{F}_{i}(q^2)=\int_{}^{}{d}^{4}k{1 \over \left[{(k-\frac{q}
{2}{)}^{2}+{M}_{i}^{2}}\right]\left[{(k+{q \over
2}{)}^{2}+{M}_{i}^{2}}\right]}\ \ \ i=u,s.
\label{eq32}
\end{equation}
Whenever the mesons lie below the threshold their mass is given by:
\begin{eqnarray}
{\Lambda }_{11}^{{}^*}(-{m}_{{f}_{0}(1370)}^{2})=0,\ \
{\Lambda }_{22}^{{}^*}(-{m}_{{f}_{0}(1710)}^{2})=0.
\label{eq33}
\end{eqnarray}
When the threshold is reached, Eq. (\ref{eq33}) has to be modified to take
into account the fact that the meson acquires a width due to its nonphysical 
decay
into $u\overline{u}$ or $s\overline{s}$. This width and its mass are 
solutions of :
\begin{equation}
{\Lambda }_{11}^{{}^*}{\left.{({q}^{2}-i\varepsilon 
)}\right|}_{{q}^{2}=-({m}_{{f}_{0}(1370)}-i{\Gamma
}_{{f}_{0}(1370)}{)}^{2}}=0,
\label{eq34}
\end{equation}
or more explicitly:
\begin{equation}
4{N}_{c}{F}_{u}\left[{-({m}_{{f}_{0}}+i\varepsilon -i{\Gamma
}_{{f}_{0}}{)}^{2}}\right]\left[{-({m}_{{f}_{0}}-i{\Gamma
}_{{f}_{0}}{)}^{2}+4{M}_{u}^{2}}\right]+{a}^{2}{\chi }_{0}^{2}{{m}_{u}
\over {M}_{u}}=0,
\label{eq35}
\end{equation}
with similar expression for ${f}_{0}(1710)$ with $u$ replaced by
 $s$.
In order to get rid of the unphysical ${\Gamma }_{{f}_{0}}$, we
perform here calculation with the approximated equations \cite{13}
\begin{equation}
4{N}_{c}\left|{{F}_{i}\left[{-({m}_{{f}_{0}}+i\varepsilon {)}^{
2}}\right]}\right|({-m}_{{f}_{0}}^{2}+4{M}_{u}^{2})+{a}^{2}{\chi
}_{0}^{2}{{m}_{i} \over {M}_{i}}=0,\ \ i=u,s.
\label{eq36}
\end{equation}
Other approximated expressions for the right-hand side of Eq. (\ref{eq35})
can be found in the literature \cite{17}. To
our knowledge, Eq.(\ref{eq35}) has never been solved exactly.

\section{Results}
\label{section5}

\begin{figure}[h]
\vbox{\psfig{file=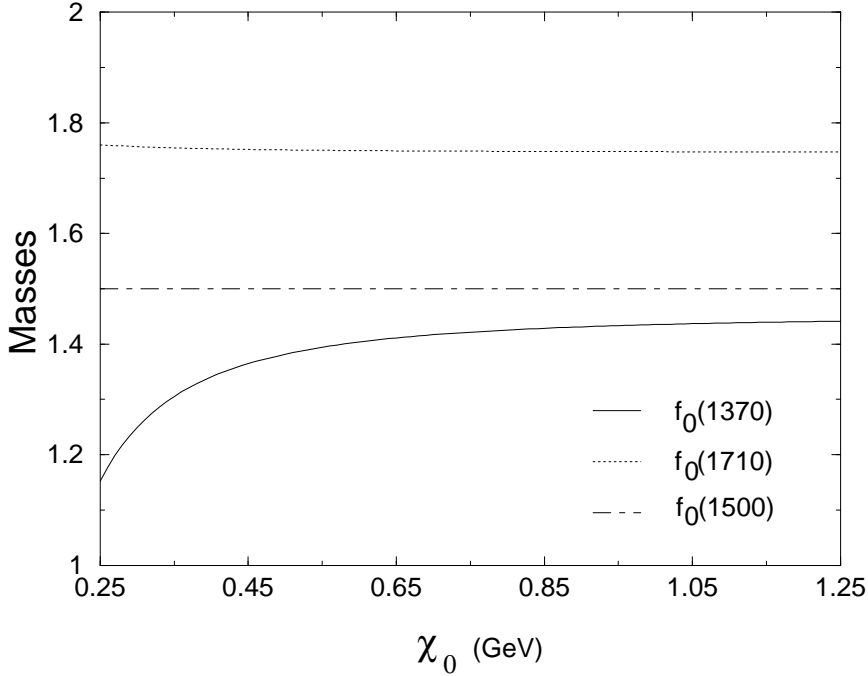,width=\widtheps}}
\caption{$\chi_0$ behavior of the mass of the three states $f_0(1370)$,
$f_0(1500)$ and $f_0(1710)$.}
\label{fig3}
\end{figure}

\begin{figure}[h]
\vbox{\psfig{file=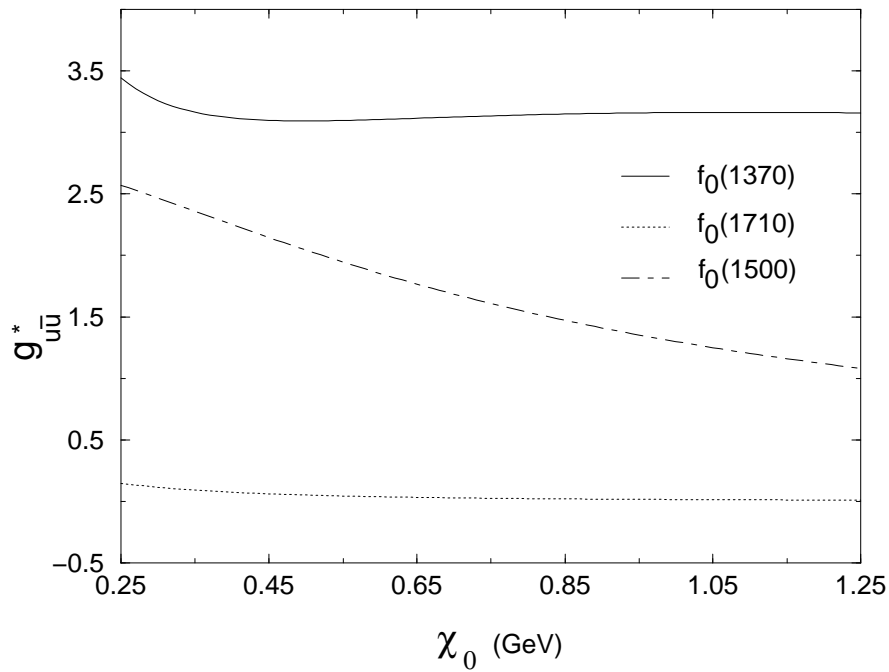,width=\widtheps}}
\caption{$\chi_0$ behavior of the coupling constant to the quark $u$,
 ${g}^{*u\overline{u}}$, of the three states $f_0(1370)$,
$f_0(1500)$ and $f_0(1710)$.}
\label{fig4}
\end{figure}

\begin{figure}[h]
\vbox{\psfig{file=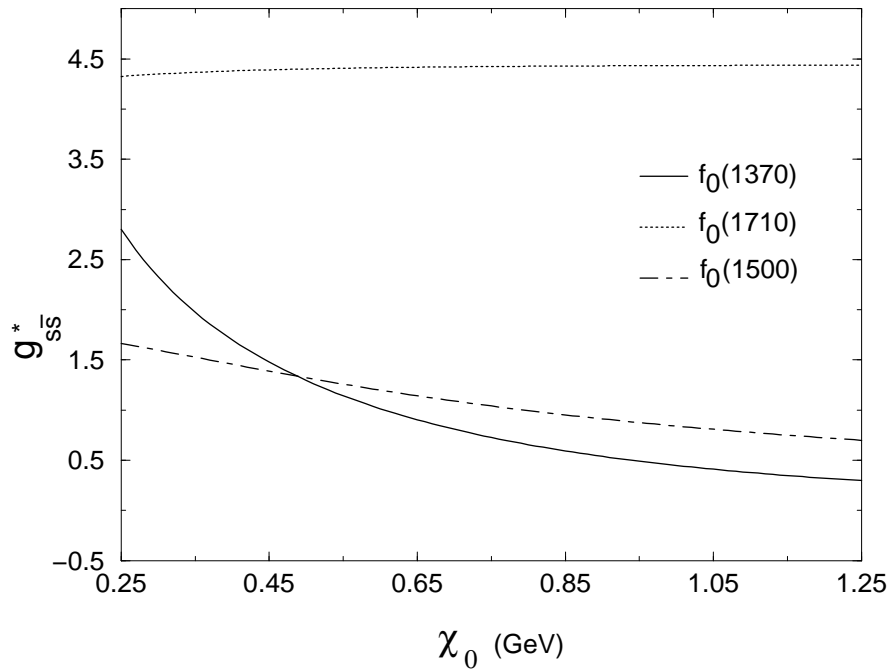,width=\widtheps}}
\caption{$\chi_0$ behavior of the coupling constant to the quark $s$,
 ${g}^{*s\overline{s}}$, of the three states $f_0(1370)$,
$f_0(1500)$ and $f_0(1710)$.}
\label{fig5}
\end{figure}

The $\chi_0$ behavior for the mass of the three $f_0$ states is
given in Fig.~\ref{fig3}. 
For large values of $\chi_0$, the states take the mass
they would have without coupling ($m_{f_{0}(1370)}= 2M_u=1450$ MeV
, $m_{f_{0}(1710)}=2M_s=1728$ MeV in the chiral limit). The mass of 
$f_{0}(1710)$ is  quite
 stable while an acceptable value for the mass of $f_{0}(1370)$ limits the 
value of $\chi_0$ to a domain between 300 and 450 MeV (see Table \ref{tab1}).
\begin{center}
\begin{table}[b]
\caption{Masses and widths in $2\gamma$ of the hybrids
$f_{0}(1370)$,
$f_{0}(1710)$ and
$f_{0}(1500)$.}
\label{tab1}
\vspace{7mm}
\begin{center}
\begin{tabular}{|c|c|c|c|c|} \hline 
& & masses(MeV)&$\Gamma$(eV)&${\Gamma}/
{\Gamma_{f_0(1710)}}$\\ \hline
&${f}_{0}(1370)$&1341 &348 &51.9\\
$\chi_{0} = 400$ MeV&${f}_{0}(1500)$&1500&102&15.2\\
&${f}_{0}(1710)$&1753 &6.7 &1\\ \hline
&${f}_{0}(1370)$&1305&617&69.7\\
$\chi_{0} =350$ MeV&${f}_{0}(1500)$&1500&112&12.7\\
&${f}_{0}(1710)$&1755&8.85&1\\ \hline
&${f}_{0}(1370)$&1250&1258&102.3\\
$\chi_{0} = 300$ MeV&${f}_{0}(1500)$&1500&123&10.0\\
&${f}_{0}(1710)$&1757&12.3&1\\ \hline
\end{tabular}
\end{center}
\end{table}
\end{center}
The same kind of behavior appears at the level of the coupling constants
of the $f_0$ states to the quarks $u$ (Fig.~\ref{fig4}, Eq.~(\ref{eq27})) and  
$s$ (Fig.~\ref{fig5}, Eq.~(\ref{eq28})). Note that Fig.~\ref{fig4} and Fig.~\ref{fig5} exhibit the modulus 
of the coupling constants. Whenever the phase factor $(m_{f_0}^{2
}-4M_i^2)$ is 
negative, the corresponding coupling constant also becomes negative in such a 
way that the two contributions $u$ and $s$ add up
 (see Eqs. (\ref{eq13})--(\ref{eq15})).
 Here again, ${g}^{*u\overline{u}}$ and
 ${g}^{*s\overline{s}}$ are stable for the $f_{0}(1710)$ with
 ${g}^{*s\overline{s}}$ an order of magnitude larger than 
 ${g}^{*u\overline{u}}$ indicating that $f_{0}(1710)$ is nearly a pure 
$s\overline{s}$ state. For large $\chi_0$, ${g}^{*u\overline{u}}_{f_0(1710)}
{\rightarrow} 0$. In contrast, the coupling constants
 ${g}^{*u\overline{u}}$ and ${g}^{*s\overline{s}}$ are of the same order 
of magnitude for $f_{0}(1370)$ and for $f_{0}(1500)$. At large $\chi_0$, 
${g}_{{f}_{0}(1370)}^{*s\overline{s}}$ goes to zero, reflecting the
fact ${f}_{0}(1370)$ is a pure $u\overline{u}$ state when the mixing is 
turned off. Finally, ${f}_{0}(1500) {\rightarrow} 2\gamma$ is forbidden in 
our 
model for large $\chi_0$, illustrated by the asymptotic 
behavior of ${g}_{{f}_{0}(1500)}^
{*u\overline{u}}$ and ${g}_{{f}_{0}(1500)}^{*s\overline{s}}$ that vanish when 
$\chi_0 {\rightarrow} \infty$. However, the decrease of the coupling 
constants of ${f}_{0}(1500)$ is slower than the others.

\begin{figure}[h]
\vbox{\psfig{file=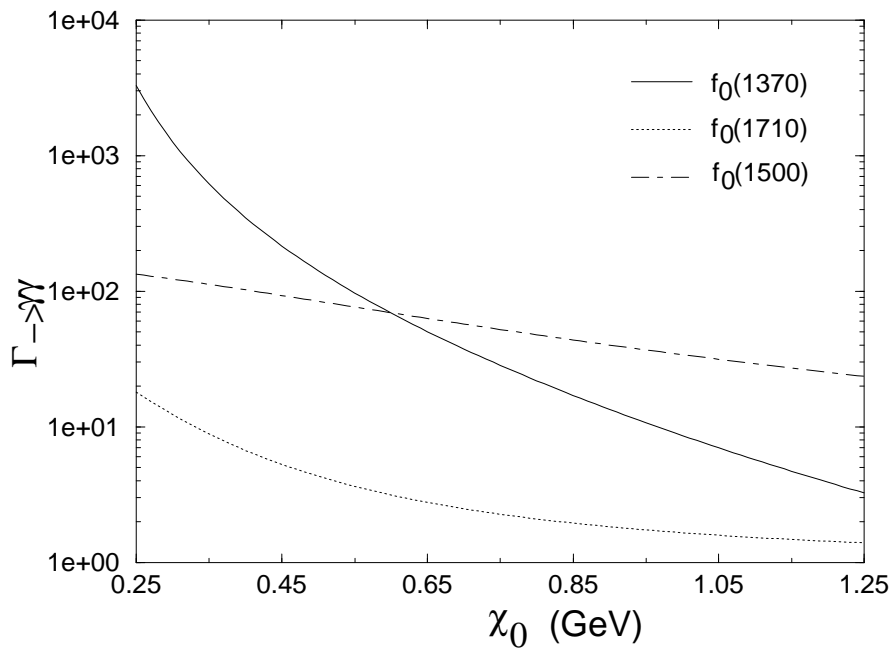,width=\widtheps}}
\caption{$\chi_0$ behavior of the width into $2\gamma$ of the three states 
$f_0(1370)$, $f_0(1500)$ and $f_0(1710)$.}
\label{fig6}
\end{figure}

The masses and the 
coupling constants so obtained allow to estimate the $2\gamma$
 decay widths (Eqs. (\ref{eq5}),(\ref{eq13})--(\ref{eq15})). Results are 
shown in
Fig.~\ref{fig6}. The three widths decrease with increasing $\chi_0$ since for $\chi_0
 \rightarrow \infty$,  at least one of the corresponding coupling constants
 vanishes. $\Gamma_{f_0(1710)}$ is always the smallest because of two 
additionnal effects:
 in the RHS of Eq. (\ref{eq14}) the first term is small due to the small 
value of ${g}_{{f}_{0}(1710)}^{*u\overline{u}}$ while the second one nearly 
vanishes due to the fact that $m_{f_0(1710)} \approx 2M_s$.  For the two other 
states, the relative amplitude of the decays depends on the value of 
$\chi_0$.

 For small $\chi_0$, $\Gamma_{f_0(1370)}>\Gamma_{f_0(1500)}$ due to four 
cumulative 
effects: 
$g_{f_0(1370)}^{*i\overline{i}} > g_{f_0(1500)}^{*i\overline{i}}$ 
and $\left|m_{f_0(1370)}^2-4M_i^2\right|>
\left|m_{f_0(1500)}^2-4M_i^2\right|$ with $i=u,s$. 
For large $\chi_0$, $m_{f_0(1370)}\approx 2M_u$ and 
${g}_{{f}_{0}(1370)}^{*s\overline{s}}$ 
becomes smaller while the RHS of Eq. (\ref{eq15}) keeps two nonvanishing terms 
up to quite
 large values of $\chi_0$. One then has $\Gamma_{f_0(1370)} < \Gamma_{f_0(1500)}$. Note that in the chiral limit, the three widths
go asymptotically to zero. (The factors $(m^2-4M^2)$, which lead to a vanishing of the decay widths of $f_0(1370)$ and $f_0(1710)$ in the pure NJL model (no mixing) in the chiral limit, are a consequence of the chiral symmetry in the scalar sector. This was already shown by Bajc et al. Ref.~\cite{BBHNR}.)

\begin{figure}[h]
\vbox{\psfig{file=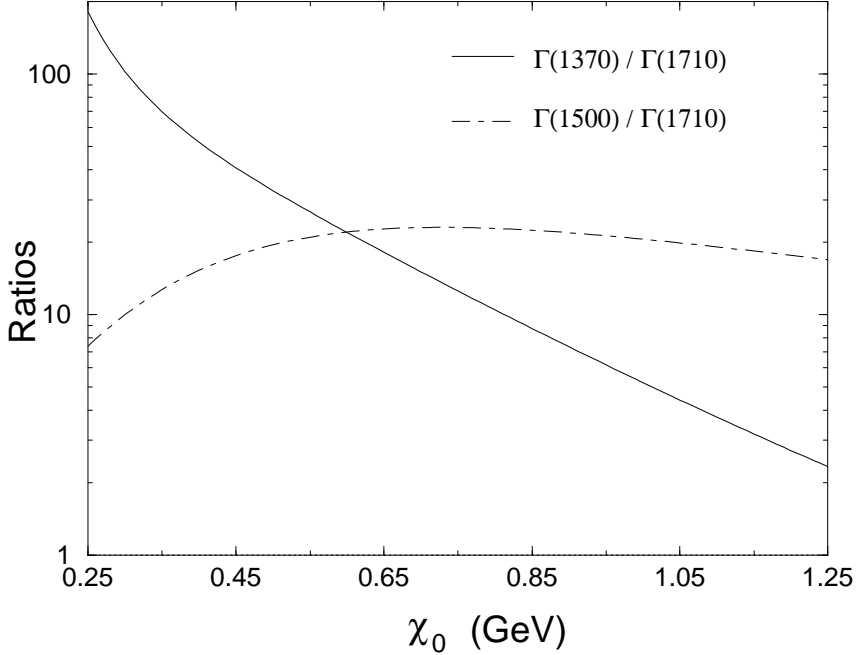,width=\widtheps}}
\caption{$\chi_0$ behavior of the ratios 
$\Gamma_{f_0(1370)}/\Gamma_{f_0(1710)}$ and 
$\Gamma_{f_0(1500)}/\Gamma_{f_0(1710)}$.}
\label{fig7}
\end{figure}

One then sees that our model provides meson and glueball widths that can vary 
from
 one to three orders of magnitude. Moreover, their relative magnitude is
 also $\chi_0$ dependent (see Fig.~\ref{fig7}). Taking $\Gamma_{f_0(1710)}$ as
 reference, $\Gamma_{f_0(1370)}/\Gamma_{f_0(1710)}$ varies
 of two
 orders of magnitude and
 $\Gamma_{f_0(1500)}/\Gamma_{f_0(1710)}$, while more stable, can however vary of a factor 5 in the considered domain of 
$\chi_0$.

 We have indicated in Table \ref{tab1} the value of the widths as well as
 their ratio to $\Gamma_{f_0(1710)}$, for three typical values of $\chi_0$
 compatible with the masses of the $f_0$ states.

Our results are in complete desagreement with the statement of Close et 
al. \cite{4} presented in the introduction: the relative strengths are not
 reproduced, even qualitatively since the width $\Gamma_{f_0(1500)}$ is
 always larger than $\Gamma_{f_0(1710)}$. Moreover, the width of $f_0(1370)$
 is always too small. To get the lower bound of the experimental value ($5.4\pm 2.3$ keV), 
one needs to consider gluon condensates as small as 270 MeV yielding 
a mass of $\approx 1200$ MeV and a still higher ratio 
$\Gamma$$/$$\Gamma_{f_0(1710)}$.

In order to compare our results with those of Ref. \cite{14bis} it seems
 appropriate to compare our ratio $\Gamma_{f_0(1370)}$$/$$\Gamma_{f_0(1500)}$
with their result for $\Gamma_{f_0(980)}$$/$$\Gamma_{f_0(1590)}$. Indeed, in 
both cases the coupling of the two mesons to the $u\overline{u}$ and $s\overline{s}$ 
channels is important. Of course our model provides an additionnal
glue content. One finds a quantity which varies from 3 to 10 while their
corresponding result lies between 6 to 11.

\section{Summary and conclusions}
\label{section6}
The model we developed in Refs. \cite{12,13} implements the QCD trace anomaly
of QCD by the introduction of a dilaton scalar field $\chi$. (Other models of this type are on the market. See for example~\cite{CER,21} and references therein.) Since the NJL 
model is not renormalizable, a cut-off must be introduced to regularize the 
diverging integrals. This cut-off breaks the scale invariance of the quark 
loop that can be restored by the replacement $\Lambda\rightarrow\Lambda\chi$.
This entails a mixing between the three scalar isoscalar fields. Put in an
 other way, one emerges with three scalar "hybrids" $f_0$ which are mixing
 of glue and $u\overline{u}$ and $s\overline{s}$ excitations. We followed the 
idea of Close et al. \cite{4} who identify these hybrids with $f_0(1370)$, 
$f_0(1500)$ and $f_0(1710)$. In our model, one has one free parameter, the 
vacuum gluon condensate $\chi_0$ that fixes the strength of the mixing. In the 
domain of $\chi_0$ (300 MeV $\leq
\chi_0 \leq 450$ MeV) which reproduces in an acceptable way the masses of the
 states, the mixing is quite large, reflected by the large values of the
 coupling constants to both types of quarks $u$ and $s$, especially for
 $f_0(1370)$ and $f_0(1500)$. The latter is identified with the glueball in
 the sense that it is the state that would yield pure glue if there was no
 mixing. In that case, it could not decay into $2\gamma$.  Here this decay is
 allowed as well as that of the two other states. The relative magnitude of
 the widths largely depends on the value of the gluon condensate. However,
 whatever $\chi_0$, the width of $f_0(1710)$ is always the smallest.
Our results are then at variance with that of Close et al. \cite{4}  according 
to which  it is the width of $f_0(1500)$ that is always the smallest. They 
also claimed, that the idea of mixing should be revised if it was found 
experimentally that $\Gamma_{f_0(1500)}$ could exceed the width of the other
states. We have however presented a model which yields totally different 
results while still including mixing and have shown that  the reduction factors $(m^2-4M^2)$ (playing almost no role for the $f_0(1500)$) are the key to understand the discrepancy between our results and those of \cite{4}. Of course, our model contains some 
drawbacks, the most important being surely the lack of confinement. However 
the statement of Ref. \cite{4} should be considered with some reserve and we 
suggest that people using model with dilatons \cite{21} coupled to scalar 
states investigate the problem. 
\vspace {10mm}

\begin{ack} We thank Isabelle Royen for numerous discussions during the completion of this work.
\end{ack}

\end{document}